\documentclass[12pt]{iopart}
\usepackage{latexsym}

\def\cR{{\cal R}}
\def\Tr{{\rm Tr}}
\def\ket#1{\mid~\!\!\!{#1}~\!\!\rangle}
\def\bra#1{\langle~\!\!{#1}~\!\!\!\mid}
\def\rogeq{\buildrel \rho \over \geq}
\def\(ro=){\buildrel \rho \over =}
\def\ro>{\buildrel \rho \over >}
\def\average#1{\langle~\!\!{#1}~\!\!\rangle}

\begin{document}\jl{1}

\title[\bf Coherence Entropy of Twin
Observables]{\bf The Role of Coherence Entropy\\
of Physical Twin Observables in
Entanglement}
\author{F Herbut\footnote[1]{E-mail:
fedorh@infosky.net}}
\address{Serbian Academy of
Sciences and Arts, Knez Mihajlova 35,
11000 Belgrade, Serbia}

\date{\today}

\begin{abstract}
The concept of physical twin observables
(PTO) for bipartite quantum states,
introduced and proved relevant for
quantum information theory in recent
work, is substantially simplified. The
relation of observable and state is
studied in detail from the point of view
of coherence entropy. Properties of this
quantity are further explored. It is
shown that, besides for pure states, also
for a class of mixed states quantum
discord (measure of entanglement) can be
expressed through the coherence entropy
of a PTO complete in relation to the
state.
\end{abstract}

\maketitle

\normalsize 
\section{\bf Introduction}
\rm This study hinges on two concepts,
that of coherence entropy and that of
twin observables. To understand what {\it
coherence} is, one starts with the lack
of it. One considers a quantum state
(density operator) $\rho$ and a discrete
observable (Hermitian operator) in
spectral form with distinct eigenvalues
$A=\sum_la_lP_l$ (possible $a_l=0$
included) because one deals with a
relative concept: observable in relation
to state. Let $B$ be another observable.
Its average in $\rho$ is
$$\average{B}= \Tr(\rho
B)=\sum_l\Tr(P_l\rho B). \eqno{(1)}$$ The
question is when is this a mixture of
separate contributions from the
eigenvalues $a_l$ of $A$, i. e.,
$$\sum_l\Tr[(P_l\rho P_l)
B]=\sum_lw_l\average{B}_l,\eqno{(2a)}$$
where $$\forall l:\quad w_l\equiv
\Tr(P_l\rho P_l)\eqno{(2b)}$$ are the
statistical weights, and, for $w_l>0$,
$$\average{B}_l \equiv \Tr[(P_l\rho
P_l/w_l)B].\eqno{(2c)}$$

One can easily convince oneself that
putting the analogous question in terms
of the classical counterparts, the answer
is: always. One says that in classical
physics there is no coherence, i. e.,
coherence is an unknown concept there.

Further, since $[A,B]=0$ implies $\forall
l:\enskip [P_l,B]=0$, it is easy to see
that (2a) is valid in this case. One says
that coherence never shows up with
respect to an observable $B$ compatible
(commuting) with the observable $A$ at
issue.

To put the above question in a more
specific form, we ask when is (1) equal
to (2a) for {\it all} observables $B$
that are {\it incompatible} with $A$. One
can prove that this is the case if and
only if $[A,\rho ]=0$, i.e., in case of
{\it compatibility of observable and
state} \cite{FHAnnPhys}.

Thus, if $A$ and $\rho$ are {\it
incompatible}, and only then, (1) is not
equal to (2a) for all $B$, and this is
called {\it coherence} of $A$ in relation
to $\rho$.

Experimentally coherence is usually
observed as {\it interference}, i. e., a
cooperative contribution of two or more
eigenevents (eigenprojectors) $P_l$ of
$A$ in the average of some observable $B$
incompatible with $A$. (Note that in case
of events, average and probability is the
same thing.)

The best known example of interference is
that on two slits. In a previous article
\cite{interf} a detailed discussion of it
is given along the lines of this
introduction (with the additional
intricacy of an evolution between passage
of the two-slit screen and arrival at the
detection screen).

The next question is how to find a
quantity that would be {\it the amount of
coherence} of $A$ in $\rho$. On intuitive
grounds one can say that it must satisfy
three requirements.

(i) It must be a function of $A$ and
$\rho$.

(ii) In case of compatibility $[A,\rho
]=0$ it must be zero, otherwise it must
be positive.

Some of the eigenevents $P_l$ of $A$ may
be compatible with $\rho$, hence part of
the average of $B$ may be expressible as
an average of separate contributions.
This part is irrelevant for coherence,
because the latter is negation of such a
mixture. Hence, the third requirement is:

(iii) The desired quantity should depend
only on those eigenevents $P_l$ of $A$
that are not compatible with $\rho$, and
not at all on those that are.

In a previous article \cite{FH02b} the
{\it amount of coherence} of $A$ in
$\rho$ was denoted by $E_C(A,\rho)$ and
defined as
$$E_C(A,\rho)\equiv S(\sum_lP_l\rho
P_l)-S(\rho),\eqno{(3)}$$ where $S(\rho)$
is the von Neumann entropy of the quantum
state $\rho$. The quantity $E_C(A,\rho)$
is the (nonnegative) entropy increase in
ideal measurement of $A$ in $\rho$. It is
called {\it the coherence entropy}. It
satisfies the first two intuitive
requirements. That it satisfies also the
third one is proved below (Theorem 2).

Physical twin observables (PTO) were
shown to be relevant \cite{FH02b} for
important questions in quantum
information theory \cite{Nielsen},
\cite{VedralRMP}. In particular, PTO can
be made use of both for defining the
quasi-classical or
subsystem-measurement-accessible part and
the purely quantum part, i. e., the
amount of entanglement or the quantum
discord \cite{Zurek} in a general
bipartite pure state.

The definition of PTO applies to two
opposite-subsystem observables (Hermitian
operators) $A_1$ and $A_2$ that stand in
a special relation to a given composite
$(1+2)$-system state (density operator)
$\rho_{12}$. The definition in
\cite{FH02b} begins with the (very
strong) requirement of {\it
compatibility} (commutation) of the
observables (operators) with the
corresponding subsystem states (reduced
density operators): $$ [A_s,\rho_s]=0,
\quad s=1,2\eqno{(4)}$$ where, of course,
the subsystem states are $\rho_s\equiv
\Tr_{s'}\rho_{12},\enskip
s,s'=1,2,\enskip s\not= s'$, and
"$\Tr_{s'}$" denotes the partial trace
over subsystem $s'$. Further, a {\it
bijection} between the {\it detectable
eigenvalues} of $A_1$ and $A_2$ is
required such that, if $P_s^i$ are the
eigenprojectors corresponding to the
$i-th$ detectable pair of eigenvalues,
$s=1,2$, connected by the bijection, then
the following, so-called {\it algebraic
condition}
$$\forall i:\quad P_1^i\rho_{12}=
P_2^i\rho_{12}\eqno{(5a)}$$ has to be
satisfied. An equivalent condition is the
measurement-theoretic one claiming that
$$\forall i:\quad P_1^i\rho_{12}P_1^i=
P_2^i\rho_{12}P_2^i.\eqno{(5b)}$$ (There
are two more equivalent conditions
\cite{FH02b} that will not be needed in
this article.)

In section 2 it is demonstrated that the
expounded definition of PTO can be
(substantially) simplified. In section 3
the concept of coherence entropy is
studied by clarifying the basic necessary
property: that the given observable
should be "discrete in relation to" the
given state. Further, some well-known
entropy inequalities are put in the form
of an equality ((16) below) and displayed
on a diagram. In section 4 the part of
the spectrum of the observable that is
actually responsible for determining the
coherence entropy is singled out. Thus,
the third intuitive requirement for
coherence entropy is shown to be valid.
In section 5 the partial order
"finer-coarser in relation to" among
observables is studied and the concept of
"complete in relation to" a given state
is investigated.

In section 6 incompatibility of
observable and state that is due
exclusively to quantum correlations is
discussed. In the final section 7 a class
of mixed states is identified in which
the coherence entropy of physical twin
observables can be viewed as constituting
the quantum discord (the entire amount of
entanglement). In the final section 8 the
main results are summed up.

\section{\bf Redundancy of
compatibility as a requirement}
Let $\rho_{12}$ be a bipartite state.
Eigenvalues of observables that have
positive probability in $\rho_{12}$ we
call detectable ones.

{\bf Theorem 1.} {\it Let $A_1$ and $A_2$
be opposite-subsystem observables, and
let there exist a bijection between all
detectable eigenvalues of $A_1$ and all
those of $A_2$ such that, upon using the
common index $i$ for the pair of
corresponding eigenvalues, the algebraic
condition (5a) is valid. Finally, let the
total probability of detectable
eigenvalues of $A_1$ and separately of
$A_2$ be $1$. Then the compatibility (4)
is a consequence.}

In proving the theorem we use the known
lemma stating the following equivalence
between two expressions of one and the
same elementary relation between an event
(projector) $P$ and a state $\rho$
expressing certainty:
$$\Tr\rho P=1\quad \Leftrightarrow
\quad P\rho =\rho \eqno{(6)}$$ (cf
\cite{FH96} if proof is wanted).

{\bf Proof} of the theorem. Let
$\{a^s_i:\forall i\}$, s=1,2, denote the
detectable eigenvalues of $A_1$ and of
$A_2$ respectively. Let, further,
$\{P_s^i:\forall i\}$ be the
corresponding eigenprojectors. (We refer
to them as to detectable ones.) Finally,
let $P_s\equiv \sum_iP^i_s\quad s=1,2$.
We write the observables in the form
$$A_s=\sum_ia^s_i
P_s^i+P_s^{\perp}A_sP_s^{\perp}, \quad
s=1,2,\eqno{(7)}$$ where $P_s^{\perp}$
denotes the orthocomplementary projector
of $P_s$.

To prove commutation of the undetectable
parts with $\rho_s$, we utilize the
second relation in (6):
$$(P_s^{\perp}A_sP_s^{\perp})\rho_s=
(P_s^{\perp}A_sP_s^{\perp})(P_s\rho_s)$$
hence
$$(P_s^{\perp}A_sP_s^{\perp})\rho_s=0=
\rho_s(P_s^{\perp}A_sP_s^{\perp}).$$ (The
last equality is due to adjoining the
preceding one.)

Commutation of the detectable parts can
be proved as follows. Making use of (5a)
and of its adjoint, one has:
$$P_s^i\rho_s=
P_s^i\Tr_{s'}\rho_{12}=\Tr_{s'}[P_s^i\rho_{12}]=
\Tr_{s'}[P_{s'}^i\rho_{12}]=\Tr_{s'}[\rho_{12}
P^i_{s'}]=\Tr_{s'}[\rho_{12}P^i_s]=(\Tr_{s'}
\rho_{12})P^i_s$$ $$=\rho_sP^i_s\quad
s,s'=1,2\quad s'\not= s.$$  Hence, in
view of (7), the validity of (4) is
established.\hfill $\Box$

Thus, one can say that two
opposite-subsystem observables $A_1$ and
$A_2$ are {\it physical twin observables}
with respect to a bipartite state
$\rho_{12}$ if the conditions of theorem
1 are valid.

It may be the case that one has physical
twin observables such that the detectable
eigenvalues that correspond to each other
via the mentioned bijection are {\it
equal} for all values of $i$, then (5a)
is replaceable by the stronger (and more
concise) algebraic relation
$$A_1\rho_{12}=A_2\rho_{12}.
\eqno{(8)}$$ {\it This relation by itself
implies all the rest of the properties of
the observables in relation to}
$\rho_{12}$ \cite{FHMV}, \cite{FHMD}. (Cf
\cite{FH02b} for the properties not
mentioned in this article.)

It is noteworthy that a comparison of (8)
and (5a) reveals that each two
corresponding detectable eigenprojectors
$P_1^i,P_2^i$ satisfy the stronger
algebraic relation (8). Hence, as proved
in \cite{FHMD}, they are compatible with
the corresponding subsystem states, i.
e., (4) is valid, {\it mutatis mutandis},
for them. Then, a glance at (7) makes it
clear that for the validity of (4) one
actually needs only to prove
compatibility of the undetectable parts
$(P_1^{\perp}A_1P_1^{\perp}),(P_2^{\perp}
A_2P_2^{\perp})$ with the corresponding
subsystem states (e. g., as done in the
proof of the theorem).

In the special case characterized by (8),
$A_1,A_2$ should be called {\it algebraic
twin observables}. They were studied in
detail in previous work
\cite{FHMV},\cite{FHMD}. There, such
observables were called simply "twin
observables". In the more recent
investigation \cite{FH02b} this practical
terminology was utilized for physical
twin observables as it should be in view
of the fact that the latter, being more
general, can be expected to have a wider
scope of application.

As it was mentioned in the Introduction,
in a recent study \cite{FH02b} the amount
of purely quantum correlations or
entanglement (or quantum discord) of an
arbitrary given pure state
$\ket{\Phi}_{12}$ was shown to be
"carried" by a specially constructed pair
of twin observables. The way how the
quantum discord is "carried" is expressed
via the notion of coherence entropy. This
result requires generalization.

In the next three sections we forget
about bipartite states and twin
observables for the time being, and we
make a precise analysis of the concept of
coherence entropy (to be able to apply it
to PTO).

\section{Observables that are discrete
in relation to a state} 

Let us rewrite the spectral form of a
given {\it discrete observable}
$A=\sum_la_lP_l$ (distinct eigenvalues)
indexing the eigenvalues and eigenevents
that are detectable in a given state
$\rho$ by $i$, and those that are
undetectable by $m$:
$$A=\sum_ia_iP_i+\sum_ma_mP_m.\eqno{(9)}$$

{\bf Lemma 1.} {\it Only the detectable
eigenevents contribute to the coherence
entropy: $$E_C(A,\rho )=
E_C\Big((\sum_ia_iP_i),\rho \Big),
\eqno{(10)}$$ and the event  $P$ defined
as $P\equiv \sum_iP_i$ is certain in
$\rho$.}

{\bf Proof.} It is easy to see (by using
the orthocomplement) that the following
claim is an equivalent form of the
relations in (6): An event $P$ is
undetectable in a state $\rho$ if and
only if $P\rho =0$. Hence, substituting
(9) in (3), (10) immediately ensues.
Further, for the same reason,
$$1=\Tr[(\sum_lP_l)\rho
]=\Tr[(\sum_iP_i)\rho ]$$. \hfill $\Box$

Lemma 1 enables one to see how widely one
can extend the set of all discrete
observables to obtain a widest set to the
elements of which the coherence entropy
concept is applicable. (This widest set
was introduced {\it ad hoc} in
\cite{FH02b}.)

 Every observable $A$ can be written in the
(partly) spectral form
$$A=\sum_ia_iP_i+P^{\perp}A
P^{\perp}\eqno{(11)}$$ (the $a_i$ are
distinct detectable eigenvalues, possible
$a_i=0$ included; $i$ enumerates all of
them). Naturally, the singled out
detectable discrete part (the first term
on the RHS) may be zero.

{\bf Definition 1.} {\it Let $A$ be an
observable and $\rho$ a state such that
the total probability of detectable
eigenvalues of $A$ in $\rho$ is one. We
say that $A$ is {\bf discrete in relation
to $\rho$}. Further, we define {\bf the
coherence entropy} $E_C(A,\rho )$ for
such an observable as}: $$E_C(A,\rho
)\equiv E_C\Big((\sum_ia_iP_i),\rho
\Big).\eqno{(12)}$$ (Note that on the RHS
we have the known coherence entropy of a
discrete observable, whereas on the LHS
we define this for $A$ that need not be
discrete in the absolute sense.)

To understand what class of observables
we are dealing with, we make some
elaboration.

{\bf Lemma 2.} {\it If a general
observable $A$ and a state $\rho$ are
given, the total probability of the
detectable eigenvalues of the former in
the latter {\bf is one if and only if},
enumerating all detectable eigenvalues by
$i$, the projector $P\quad (\equiv
\sum_iP_i)$ projects onto a subspace that
{\bf contains} the range of $\rho$.}

If $Q$ denotes the range projector of
$\rho$, then the algebraic form of the
(geometric) characteristic condition in
the lemma is
$$PQ=Q.\eqno{(13)}$$ This is a further
known characteristic condition (cf (6))
for an event being certain in a state.
(For the reader's convenience, it is
proved in Appendix 1.)

{\bf Lemma 3.} {\it If $[A,\rho ]=0$,
then also $[A,Q]=0$, and the reducee $A'$
of $A$ in the range $\cR(Q)$ is {\bf
discrete} (in the absolute sense). Its
eigenvalues are precisely the detectable
eigenvalues $\{a_i:\forall i\}$ of $A$.
The observable $A$ is discrete in
relation to $\rho$.}

{\bf Proof.} Since $A$ commutes with all
eigenprojectors of $\rho$, and
$Q=\sum_kQ_k$ (sum of all eigenprojectors
of $\rho$ corresponding to positive
eigenvalues), also $[A,Q]=0$ is valid. On
account of $\forall k:\enskip [A,Q_k]=0$,
$A$ reduces in each eigensubspace
$\cR(Q_k)$ of $\rho$. Since these are
finite dimensional (the positive
eigenvalues of $\rho$ add up to $1$),
$A'$ is discrete. Since by definition
$\forall i:\enskip 0<\Tr(P_i\rho
)=\Tr[(P_iQ)\rho ]$, one has $P_iQ\not=
0$. Hence, the reducee of $P_iQ$ in
$\cR(Q)$ is a nonzero projector. It is
the eigenprojector of $A'$ corresponding
to the eigenvalue $a_i$. Thus, each
detectable eigenvalue of $A$ is an
eigenvalue of $A'$. On the other hand,
each eigenvalue of $A'$ (it is, of
course, also an eigenvalue of $A$) is
detectable because the reducee of $\rho$
in $\cR(Q)$ is nonsingular (see Appendix
2 for this implication). Finally, the sum
$\sum_i\Tr P_i\rho =\sum_i\Tr[(P_iQ)\rho
]$ must be $1$ because, designating by
prim the reducees in $\cR(Q)$, one has
$\Tr[(P_iQ)\rho ]=\Tr[(P_iQ)'\rho' ]$,
and $\sum_i\Tr[(P_iQ)'\rho' ]=\Tr\rho' =
\Tr\rho =1$. Hence, $A$ is discrete in
relation to $\rho$ as claimed.\hfill
$\Box$

Henceforth and throughout this article
when a state is given then by an
observable we mean one that is discrete
in relation to the state (or otherwise
we'll say that we have a more general
observable).

In general, the spectral part $P^{\perp}A
P^{\perp}$ (see (11)) may contain a
continuous spectrum if the null space of
$\rho$ is infinite dimensional. (If the
null space is finite dimensional, then
the observable must be discrete in the
absolute sense). In the rest of this
section we expound a basic relation of
the coherence entropy $E_C(A,\rho)$ to
its "neighboring" entropies. It is an
immediate consequence of the following
fundamental inequalities. (They have
become of classical value in quantum
entropy theory.) Let $\forall i:\quad
p_i\equiv \Tr(P_i\rho )$. Then
\cite{Lindblad}
$$\sum_ip_iS(P_i\rho P_i/p_i)\leq S(\rho
)\leq S(\sum_iP_i\rho P_i).
\eqno{(14a,b)}$$ One has equality in the
first inequality if and only if $\forall
i:\enskip S(P_i\rho P_i/p_i)=S(\rho )$.
The second inequality reduces to an
equality if and only if one has
compatibility: $\forall i:\enskip
[P_i,\rho ]=0$.

Since $\sum_iP_i\rho P_i$ is an
orthogonal mixture, the mixing property
of entropy \cite{Wehrl} applies:
$$S(\sum_iP_i\rho P_i)=H(p_i)+\sum_ip_i
S(P_i\rho P_i/p_i)\eqno{(15a)}$$ where
$$H(p_i)\equiv -\sum_ip_ilnp_i
\eqno{(15b)}$$ is the Shannon entropy of
the probability distribution
$\{p_i:\forall i\}$. It is often called
the mixing entropy. But for our purposes
we view it as the {\it entropy of the
observable} $A$ in $\rho$, and denote it
by $S(A,\rho )$. This is a well known
concept (cf e. g. \cite{Grabowski78}). It
equals the amount of information that on
can gain about $A$ when measuring it in
$\rho$.

We can thus rewrite (15a) as follows:
$$S(A,\rho )=E_C(A,\rho
)+\Big(S(\rho )-\sum_ip_iS(P_i\rho
P_i/p_i)\Big).\eqno{(16)}$$ It is
noteworthy that $E_C(A,\rho )$ is
nonnegative on account of (14b), and the
second term on the RHS of (16) is also
nonnegative due to (14a). (Relation (16)
is the same as relation (20) in
\cite{FH02b}.)

 It is obvious from
(16) that $$E_C(A,\rho )\leq S(A,\rho ).
\eqno{(17)}$$ In other words, when
measuring $A$ in $\rho$, the amount of
the obtainable information on $A$ is
larger or equal to the entropic "price"
that one has to pay (for measuring an
incompatible observable).

In case of compatibility $[A,\rho ]=0$,
(16) reduces to $$S(A,\rho )=S(\rho
)-\sum_ip_iS(P_i\rho P_i/p_i).$$ Since
this is the mixing property of entropy
applied to the orthogonal decomposition
$\rho =\sum_iP_i\rho P_i$, one has
$S(A,\rho )=0$ if and only if $\forall
i:\enskip P_i\rho P_i/p_i=\rho$ (a
strengthened from of the characteristic
conditions for equality in (14a)).

Equality (16) is displayed on the
following self-explanatory {\it entropy
level diagram} with vertical distances
representing the entropy quantities at
issue.

$$\mbox{\bf Diagram}$$

$$\mbox{\rule{5cm}{.5mm}}S(\sum_iP_i\rho
P_i)
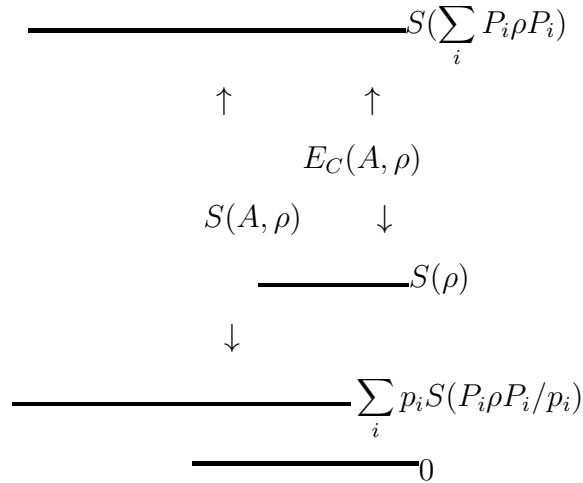
$$
$$\uparrow \qquad \qquad \uparrow$$
$$\qquad \quad \quad E_C(A,\rho )$$
$$S(A,\rho)  \qquad \downarrow$$
$$\qquad \qquad
\mbox{\rule{2cm}{.5mm}}S(\rho )$$
$$\downarrow\qquad \qquad$$
$$\mbox{\rule{4.5cm}{.5mm}}
\sum_ip_iS(P_i\rho P_i/p_i)$$
$$\quad
\mbox{\rule[2mm]{3cm}{.5mm}}0$$

\section{The spectral part of the
observable that determines the
coherence entropy}
 Let a state $\rho$ be given and let us consider
the detectable eigenprojectors $P_i$ of a
given observable $A$. Some of these may
commute with $\rho$. Following the
terminology used in \cite{FH02a}, we call
the eigenprojectors that do not commute
with $\rho$ {\it weak} ones; those that
do commute, we call {\it strong} ones.
For further analysis we enumerate the
weak eigenprojectors by $j$, and the
strong ones by $k$. Further, we write the
spectral form of $A$ as consisting, in
general, of a {\it weak component
observable} $A_w$, a {\it strong
component observable} $A_{st}$ and the
irrelevant part with respect to $\rho$:
$$A=A_w+A_{st}+
P^{\perp}AP^{\perp} \eqno{(18a)}$$ where
$$A_w\equiv
\sum_ja_jP_j\eqno{(18b)}$$ and
$$A_{st}\equiv \sum_ka_kP_k.
\eqno{(18c)}$$

Let us further define the {\it weak
probability} $p_w$, and, for the case
when it is nonzero, the {\it weak
component state} $\rho_w$:
$$p_w\equiv \Tr[(\sum_jP_j)\rho ]
\quad \rho_w\equiv (\sum_jP_j) \rho
(\sum_jP_j)\Big/p_w. \eqno{(19a,b)}$$
Then we have  the following {\it
orthogonal decomposition} of the state:
$$\rho =p_w
\rho_w+(1-p_w)\rho_{st}\eqno{(20)}$$
where the {\it strong component state} is
given by the {\it orthogonal mixture of
states}:
$$\rho_{st}\equiv
\sum_k\{[p_k/(1-p_w)]\rho_k\}\eqno{(21a)}$$
and
$$\forall k:\qquad p_k\equiv \Tr(P_k
\rho ),\qquad \rho_k\equiv P_k\rho
P_k\Big/p_k= P_k\rho
\Big/p_k.\eqno{(21b,c)}$$

{\bf Definition 2}. {\it In the three
cases when $p_w=0$, $p_w=1$, and
$0<p_w<1$ we say that the observable $A$
is strong, weak, and intermediary
respectively regarding $\rho$.}

If $\rho$ is a pure state, then every
observable is weak with respect to it
(with the exception of a constant times a
projector that does not change this state
vector). If, on the other hand, one has
$[A,\rho ]=0$, which is equivalent to the
property that all detectable
eigenprojectors of $A$ are strong, then
$A$ is strong.

{\bf Remark 1.} {\it The eigenprojectors
$P_k$ are called strong because each of
them "cuts" a separate component state
$\rho_k$ "out of" $\rho$ (cf (21c)),
whereas the weak eigenprojectors $P_j$
"cut out" a component state only in
cooperation of all of them (cf (19b)).}

{\bf Theorem 2.} {\it The coherence
entropy of $A$ is equal to the coherence
entropy of the weak component observable
in the weak component state multiplied by
the weak probability}:
$$E_C(A,\rho )=p_wE_C(A_w,\rho_w).\eqno{(22)}$$

{\bf Proof}. Definition (12) with (3),
and (20) with definitions (19a,b) and
(21a-c) imply:
$$LHS(22)=S\Big(p_w\sum_i(P_i\rho_wP_i)
+(1-p_w)\sum_i(P_i\rho_{st}P_i)\Big)
-S\Big(p_w\rho_w+(1-p_w)\rho_{st}\Big)$$
$$=S\Big(p_w\sum_j(P_j\rho_wP_j)+
(1-p_w)\rho_{st}\Big)-S\Big(p_w\rho_w+
(1-p_w)\rho_{st}\Big).$$ Since both
entropies are taken of orthogonal state
decompositions, one can apply the mixing
property of entropy, and obtain
$$LHS(22)=[H(p_w)+p_wS(\sum_jP_j\rho_wP_j)
+(1-p_w)S(\rho_{st})]-[H(p_w)+p_wS(\rho_w)
+(1-p_w)S(\rho_{st}]$$ ($H(p_w)$ being
the Shannon entropy of the probability
distribution $\{p_w,(1-p_w)\}$). After
cancellations the RHS(22) is obtained.
\hfill $\Box$

\section{Finer observables and complete
ones in relation to a state} 

Now we turn to a special relation of two
observables with respect to a given
state.

{\bf Definition 3.} {\it Let $\rho$ be a
given state and let $A$ and $A'$ be two
observables. Let further the detectable
eigenprojectors of $A'$ further decompose
those of $A$:
$$\forall i:\quad P_i=
\sum_{i'}P_{i,i'},\eqno{(23)}$$ where
$\{P_{i,i'}:\forall i,i'\}$ are
eigenprojectors of $A'$ corresponding to
its distinct eigenvalues
$\{a'_{i,i'}:\forall i,i'\}$. Then $A'$
is finer than or a refinement of $A$ and
the latter is coarser than or a
coarsening of the former in relation to
$\rho$ and we write $A'\rogeq A$. If at
least one of the sums in (23) is
nontrivial in the sense that it has at
least two detectable terms on the RHS,
then $A'$ is strictly finer than $A$ etc.
in relation to $\rho$, and we write
$A'\ro> A$. Otherwise, $A'$ and $A$ are
equal in relation to $\rho$, and we write
$A'\(ro=) A$.}

{\bf Lemma 4.} {\it If two observables
$A'$ and $A$ are such that the former is
a refinement of the latter in relation to
the state $\rho$, then the entropy of the
latter in $\rho$ does not exceed that of
the former:
$$A'\rogeq A\quad \Rightarrow \quad
S(A',\rho )\geq S(A,\rho ).\eqno{(24)}$$
The entropies are equal if and only if
 $A'\(ro=) A$.}

{\bf Proof.} Evidently, (23) implies
$\forall i:\quad p_i=\sum_{i'}p_{i,i'}$,
where $p_i\equiv \Tr(P_i\rho )$ and
$p_{i,i'}\equiv \Tr(P_{i,i'}\rho )$. One
can write
$$p_{i,i'}=\sum_mp_m[\delta_{m,i}(p_{i,i'}
/p_i)]$$ where $m$ takes on the same
values as $i$, and $p_m\equiv \Tr(P_m\rho
)$. Denoting by $p^{(m)}_{i,i'}$ the
probability distributions
$[\delta_{m,i}(p_{i,i'} /p_i)]$ on the
set of all pairs $(i,i')$, one has
disjointness
$p^{(m)}_{i,i'}p^{(m')}_{i,i'}=\delta_{m,m'}
(p^{(m)}_{i,i'})^2$. Hence, we are
dealing with a disjoint decomposition of
the probability distribution $p_{i,i'}$
(the classical discrete counterpart of an
orthogonal decomposition of a quantum
state), and we can apply the mixing
property resulting in the following
relation between Shannon entropies (cf
(15b)):
$$H(p_{i,i'})=
H(p_i)+\sum_mp_mH(p_{i,i'}^{(m)}).$$ On
account of the definition of the entropy
of an observable in a state (see the text
beneath (15b)), the last relation can be
written as
$$S(A',\rho )= S(A,\rho
)+\sum_mp_mH(p_{i,i'}^{(m)}).
\eqno{(25)}$$ The nonnegativity of the
last term bears out the first claim. If
$A'\ro> A$, at least one of the sums in
(23) is nontrivial, e. g., for $i=m$.
Then at least two pairs of indices
$(m,i'),(m,i'')\enskip i'\not= i''$
enumerate detectable eigenvalues of $A'$,
hence, also the decomposition of the
corresponding probability is nontrivial,
and $H(p_{i,i'}^{(m)})$ is positive. This
proves the second claim.\hfill $\Box$

{\bf Theorem 3.} {\it If $A'$ is a
refinement of $A$ in relation to a given
state $\rho$, then the coherence entropy
of the latter does not exceed that of the
former in this state:
$$E_C(A',\rho )\geq E_C(A,\rho ).
\eqno{(26)}$$ The two entropies are equal
if and only if the observable $A'$ is
compatible with the state $\sum_iP_i\rho
P_i$, i. e.,
$$[A',\sum_iP_i\rho P_i]=0.\eqno{(27a)}$$
The coherence entropy of $A'$ is strictly
larger than that of $A$ (both in $\rho$)
if and only if there exists a nontrivial
sum in (23), and one has for the
corresponding value of $i$:
$$\exists i'\not= i'':\quad P_{i,i'}\rho
P_{i,i''}\not= 0.\eqno{(27b)}$$.}

{\bf Proof.} Measurement of $A'$ in an
ideal way changes both $\rho$ and the
state $\sum_iP_i\rho P_i$ into one and
the same state $\sum_i\sum_{i'}
P_{i,i'}\rho P_{i,i'}$. Hence,
$S(\sum_i\sum_{i'}P_{i,i'}\rho
P_{i,i'})\geq S(\sum_iP_i\rho P_i)$ (cf
(14b)). Inequality (26) then follows (cf
(12) and (3)). Criterion (27a) is a
consequence of that for equality in
(14b). The last claim follows from the
facts that (27a) is equivalent to
$\sum_i\sum_{i'}P_{i,i'}\rho
P_{i,i'}=\sum_iP_i\rho P_i$, and that
(27b) is the negation of this.\hfill
$\Box$

Now we turn to exploring the last term in
(16) with respect to comparison of $A$
and a finer obseravable $A'$ in relation
to $\rho$. Though the probabilities $p_i$
have to be positive due to the definition
of the indices $i$, this is not the case
with $p_{i,i'}$. If this probability is
zero, the corresponding state and its
entropy are not defined. But, for
simplicity, we assume, as it is usually
done in such cases, that
$p_{i,i'}S(P_{i,i'}\rho
P_{i,i'}/p_{i,i'})$ is simply zero.

{\bf Lemma 5.} {\it If $A'$ is a
refinement of $A$ in $\rho$, then the
entropy decrease, i. e., the second term
on the RHS of (16), corresponding to $A'$
is larger than or equal to that
corresponding to $A$: $$\{S(\rho
)-\sum_i\sum_{i'} [p_{i,i'}S(P_{i,i'}\rho
P_{i,i'}/p_{i,i'})]\}\geq \{S(\rho )-
\sum_i[p_iS(P_i\rho
P_i/p_i)]\}.\eqno{(28)}$$ One has
equality if and only if} $$\forall
i,i',\enskip p_{i,i'}>0:\quad
S(P_{i,i'}\rho P_{i,i'}/p_{i,i'})=
S(P_i\rho P_i/p_i).$$

This condition is satisfied when
$A'\(ro=) A$. But it may be valid also
for $A'\ro> A$.

{\bf Proof.} One has
$$LHS(28)-RHS(28)=$$ $$
\sum_i\Bigg[p_i\Big[S(P_i\rho
P_i/p_i)-\sum_{i'}\Bigg(
(p_{i,i'}/p_i)S[P_{i,i'}(P_i\rho
P_i/p_i)P_{i,i'}\Big/(p_{i,i'}/p_i)]
\Bigg)\Big]\Bigg]$$ since
$P_{i,i'}=P_{i,i'}P_i$ (cf (23)). Each
term in the sum "$\sum_i$" is nonnegative
on account of (14a). The last claim
follows from the equality conditions in
(14a). \hfill $\Box$

{\bf Remark 2.} {\it If $A'\ro> A$ but
$[A',\sum_iP_i\rho P_i]=0$, then
$S(A',\rho )>S(A,\rho )$, but
$E_C(A',\rho )= E_C(A,\rho )$. In this
case the claimed criterion for equality
in Lemma 5, i. e., for lack of
enlargement in the entropy decrease, is
not valid. Namely, $\forall i:\enskip
P_i\rho P_i/p_i=\sum_{i'}[
(p_{i,i'}/p_i)(P_{i,i'}\rho P_{i,i'}/
p_{i,i'})]$, and the average entropy in a
mixture is always less than that of the
mixture itself unless the mixture is
trivial; but this is not the case for at
least one value of $i$.}

{\bf Definition 4.} {\it If $A'\rogeq A$
implies $A'\(ro=) A$, i. e., if $A$ does
not have a nontrivial refinement in
relation to the given state $\rho$, then
we say that $A$ is {\bf complete in
relation to} $\rho$.}

{\bf Remark 3.} {\it Evidently, $A$ is
complete in relation to $\rho$ if and
only if further decomposition is not
possible, i. e., for each detectable
eigenprojector $P_i$ of $A$ there exists
a state vector $\ket{i}$ such that:}
$$P_i\rho P_i/[\Tr(P_i\rho )]=
\ket{i}\bra{i}.\eqno{(29)}$$

{\bf Lemma 6.} {\it If an observable $A$
and a state $\rho$ in relation to which
the former is discrete are {\bf
compatible}, i. e., $[A,\rho ]=0$, and if
$Q$ denotes the range projector of
$\rho$, then the observable is {\bf
complete} in relation to the state {\bf
if and only if} for each value of $i$
there exists a state $\ket{i}$ such that
$P_iQ= \ket{i}\bra{i}$.}

{\bf Proof.} On account of the
commutation, one can choose a common
eigenbasis of $A$ and $\rho$. Let us
write its subbasis spanning $\cR(Q)$ as
$\{\ket{i,k}:\forall i,\forall k\}$ where
$\forall i:\enskip
P_i\ket{i,k}=\ket{i,k}$. Denoting the
corresponding eigenvalues of $\rho$ by
$\{r_{i,k}:\forall i,\forall k\}$, we can
write the spectral form $P_i\rho P_i=\rho
(P_iQ) =\sum_kr_{i,k}\ket{i,k}
\bra{i,k}$. It is now obvious that one
has completeness (cf (29)) if and only if
for each value of $i$ there is only one
value of $k$. Then $\ket{i}\equiv
\ket{i,k}$.\hfill $\Box$

{\bf Remark 4.} {\it Since in case of
compatibility of $A$ and $\rho$ one has
$\cR(P_iQ)= \cR(P_i)\cap \cR(Q)$, the
projector $\ket{i}\bra{i}$ is the reducee
$(P_iQ)'$ of $PQ$ in $\cR(Q)$. Therefore,
an equivalent form of the criterion in
Lemma 6 is completeness (in the absolute
sense) of the reducee $A'$ of $A$ in
$\cR(Q)$.}

\section{Correlations incompatibility}
Now we turn to bipartite states.

{\bf Lemma 7.} {\it A subsystem
observable $A_1\otimes 1$ is {\bf
discrete} in relation to a state
$\rho_{12}$ {\bf if and only if} so is
$A_1$ in relation to $\rho_1$
$\Big(\equiv \Tr_2\rho_{12}\Big)$.}

{\bf Proof} follows immediately  from
$\forall i:\enskip \Tr
(P_1^i\rho_{12})=\Tr
(P_1^i\rho_1)$.\hfill $\Box$

Let us introduce a general concept.

{\bf Definition 5.} {\it When a
first-subsystem observable $A_1$ and a
bipartite state $\rho_{12}$ are related
so that $[A_1,\rho_s]=0,\quad s=1,2$ and
$[A_1,\rho_{12}]\not= 0$, and the
observable is discrete in relation to
$\rho_1$, then we say that we have a case
of {\bf correlations incompatibility}.
The same term will be used for the
symmetric case with a second-subsystem
observable $A_2$.}

Intuitively one expects that in this case
the very incompatibility
$[A_s,\rho_{12}]\not= 0$, being due only
to the correlations in $\rho_{12}$,
should, through its amount, play a role
in understanding the correlations
contained in $\rho_{12}$, $s=1$ or $2$.
Therefore, we investigate the relation
between the coherence entropy
$E_C(A_1,\rho_{12})$ and the von Neumann
mutual information $I(\rho_{12})\equiv
S(\rho_1)+S(\rho_2)-S(\rho_{12})$.

{\bf Theorem 4.} {\it In the case of {\bf
correlations incompatibility}, one has
$$I(\rho_{12})=E_C(A_s,\rho_{12})
+I(\sum_iP_s^i\rho_{12}P_s^i), \quad
s=1\enskip \mbox{or} \enskip
2,\eqno{(30)}$$ i. e., {\bf the coherence
entropy is a term in the von Neumann
mutual information} of $\rho_{12}$
(together with another possible
nonnegative term).}

{\bf Proof}. One has always (in obvious
notation) $S_{12}= S_1-I_{12}+S_2$. We
assume that $s=1$. In our case
  $\sum_iP_1^i\rho_1P_1^i=\rho_1$, hence
$$E_C(A_1,\rho_{12})\equiv
S(\sum_iP^i_1\rho_{12}P^i_1)-S(\rho_{12})
$$
$$=\Big[S(\sum_iP^i_1\rho_1P^i_1)-I(\sum_i
P^i_1\rho_{12}P^i_1)+S(\Tr_1[(\sum_iP_1^i
)\rho_{12}])\Big]-
[S(\rho_1)-I(\rho_{12})+S(\rho_2)]$$ $$=
I(\rho_{12})-
I(\sum_iP_1^i\rho_{12}P_1^i)$$ because,
$\sum_iP_1^i$ being a certain event in
$\rho_{12}$, one has
$(\sum_iP_1^i)\rho_{12}=\rho_{12}$ (cf
(6)). If $s=2$, the proof is symmetrical.
\hfill $\Box$

Thus, the above intuitive expectation
turned out to be correct. But it is not
clear if the coherence entropy at issue
belongs to the amount of quasi-classical
correlations or to that of entanglement
in $\rho_{12}$. The latter seems more
likely because coherence and
incompatibility are unknown in classical
physics.

\section{Back to physical twin observables}
Let us start by generalizing the
mentioned result from \cite{FH02b} that a
tailor-made pair of physical twin
observables (PTO) $A_1,A_2$ for an
arbitrary given bipartite pure state
$\rho_{12}\equiv
\ket{\Phi}_{12}\bra{\Phi}_{12}$ "carries"
the amount of entanglement (quantum
discord) via the coherence entropy of
$A_1$ (or of $A_2$) in $\ket{\Phi}_{12}$.

{\bf Lemma 8.} {\it If $A_1,A_2$ are PTO
for $\ket{\Phi}_{12}$ and $A_1$ is
complete in relation to $\rho_1$, the
state (reduced density operator) of
subsystem $1$, then
$E_C(A_1,\ket{\Phi}_{12})=S(\rho_1)$ and
thus these PTO "carry" the entire amount
of entanglement, i. e., the quantum
discord in} $\ket{\Phi}_{12}$.

{\bf Remark 5.} {\it Bearing in mind that
for PTO $A_1$ and $\rho_1$ are compatible
(cf Theorem 1) $A_1$ is complete in
relation to $\rho_1$ if and only if its
reducee $A_1'$ in $\cR(Q_1)$ ($Q_1$ being
the range projector of $\rho_1$) is
complete in the absolute sense (cf Remark
4).}

{\bf Proof} of Lemma 8 is obtained by
reducing this case to the "tailor-made"
one from \cite{FH02b}. Namely, the
commutation $[A_1,\rho_1]=0$ enables one
to choose the eigen-sub-basis of $\rho_1$
that spans its range as an
eigen-sub-basis also of $A_1$. Expansion
of $\ket{\Phi}_{12}$ in this basis then
gives a Schmidt biorthogonal form, and
$A_1,A_2$ have (the "tailor-made")
spectral forms like in
\cite{FH02b}.\hfill $\Box$

We go now to a {\it general bipartite
state} $\rho_{12}$.

As it is well known, there is no entropy
increase in ideal measurement of the
observable at issue if and only if the
observable and state are compatible.
Since for PTO (4) is valid, we have a
case of correlations incompatibility if
$A_1$ has a nonzero weak component in
relation to $\rho_{12}$.

Further investigation requires a general
result concerning so-called biorthogonal
mixtures of bipartite states. Let us
define this concept.

{\bf Definition 6.} {\it Let
$\{P_1^k:\forall k\}$ and
$\{Q_2^k:\forall k\}$ be any sets of {\bf
orthogonal} projectors for the first and
the second subsystem respectively with
common enumeration, and let
$\sum_kp_k\rho_{12}^k$ be a mixture such
that $\forall k:\quad \rho_{12}^k=
P_1^k\rho_{12}^kQ_2^k$. Then the {\bf
mixture} is said to be {\bf
biorthogonal}.}

Now we can prove a general result. It is
the analogue of the mixing property of
entropy. It can be called {\bf mixing
property of von Neumann mutual
information}.

{\bf Lemma 9.} {\it Let $\rho_{12}=
\sum_kp_k\rho_{12}^k$ be a {\bf
biorthogonal mixture}. Then,
$$I(\rho_{12})=H(p_k)+\sum_kp_kI(\rho_{12}^k).
\eqno{(31)}$$ Thus the von Neumann mutual
information is the sum of the mixing
entropy and the average von Neumann
mutual information of the component
states in the mixture.}

{\bf Proof.} By definition $I(\rho_{12})=
S(\rho_1)+S(\rho_2)-S(\rho_{12})$. The
mixture entails
$\rho_s=\sum_kp_k\rho_s^k, \quad s=1,2$,
where $\forall k:\quad \rho_s^k\equiv
\Tr_{s'}\rho_{12}^k,\quad s,s'=1,2,\quad
s\not= s'$. Since, besides the composite
mixture, also both subsystem mixtures are
orthogonal, one has three mixing
properties of entropy:
$S(\rho_{12})=H(p_k)+\sum_kp_kS(\rho_{12}^k)$,
$S(\rho_s)=H(p_k)+\sum_kp_kS(\rho_s^k),
\quad s=1,2$. Substituting the three
entropy decompositions in the above
definition of mutual information, one
obtains the claimed relation (31).\hfill
$\Box$

{\bf Theorem 5.} {\it If $A_1$ and $A_2$
are {\bf physical twin observables} for
$\rho_{12}$, then
$$I(\rho_{12})=S(A_s,\rho_{12})+
E_C(A_s,\rho_{12})+\sum_ip_i
I(\rho_{12}^i)\quad s=1,2\eqno{(32a)}$$
where $$\forall i:\quad \rho_{12}^i\equiv
P_1^i\rho_{12}P_1^i/p_i=P_2^i\rho_{12}P_2^i/p_i
\eqno{(32b)}$$ and $\sum_ia_iP_s^i$ is
the detectable part of $A_s$, $\forall i:
\enskip p_i\equiv \Tr(\rho_{12}P_s^i)$,
$s=1,2$. Thus, both the entropy and the
coherence entropy of $A_s$ in $\rho_{12}$
are parts of the von Neumann mutual
information. Besides, $S(A_1,\rho_{12})=
S(A_2,\rho_{12})$, and
$E_C(A_1,\rho_{12})=E_C(A_2,\rho_{12})$.}

{\bf Proof.} The last term in relation
(30) is the mutual information of a
biorthogonal mixture (cf Definition 6 and
(5b)). Applying the mixing property of
mutual information (Lemma 9) to it, the
claimed relation (32a) is immediately
derived because
$H(p_i)=S(A_s,\rho_{12})$. The entropies
of $A_1$ and $A_2$ are equal because they
are both given by $H(p_i)$. Besides, also
the coherence entropies coincide due to
(5b) (rewritten in (32b)) (cf (12) and
(3)). \hfill $\Box$

{\bf Theorem 6.} {\it If $A_1$ and $A_2$
are PTO for $\rho_{12}$ and $A_s$ are
{\bf complete} in relation to $\rho_s$,
$s=1,2$, then
$$I(\rho_{12})=S(A_s,\rho_{12})+E_C(A_s,
\rho_{12})\quad s=1,2.\eqno{(33)}$$

The entropy $S(A_s,\rho_{12})$ of $A_s$
in $\rho_{12}$ is the quasi-classical or
subsystem-measurement accessible part,
and the coherence entropy
$E_C(A_s,\rho_{12})$ is {\bf the quantum
discord}, i.e., the amount of
entanglement, in the von Neumann mutual
information in $\rho_{12}$.}

{\bf Proof.} By assumption we have
relative completeness, i. e., $\forall
i:\enskip Q_sP_s^i=\ket{i}_s\bra{i}_s$,
where $Q_s$ is the range projector of
$\rho_s$ (cf Lemma 3 and Remark 4),
$s=1,2$. (Naturally, $[Q_s,P_s^i]=0$ is a
consequence of (4).) Relations (32b)
imply
$$\rho_s^i\equiv \Tr_{s'}\rho_{12}^i
=(\ket{i}_s\bra{i}_s\rho_s\ket{i}_s
\bra{i}_s)/p_i =\ket{i}_s\bra{i}_s,\quad
s,s'=1,2 \quad s\not= s'.$$ Thus,
$\forall i:\enskip
\rho_{12}^i=\ket{i}_1\bra{i}_1 \otimes
\ket{i}_2\bra{i}_2$, hence
$I(\rho_{12}^i)=0$, and (32a) reduces to
(33) as claimed.

To obtain the quasi-classical part
$I_{qcl}$ of $I(\rho_{12})$, we first
evaluate the subsystem entropies, which
are upper bounds for the former (cf
(4a,b) in \cite{FH02b}). Since $[\rho_s,
A_s]=0$ (cf (4)), $\rho_s$ and $A_s$ have
a common eigenbasis in $\cR(Q_s)$,
$s=1,2$. Further, the reducees $A'_s$ in
$\cR(Q_s)$ are complete (cf Remark 4),
hence the common eigenbasis in $\cR(Q_s)$
is the eigenbasis $\{\ket{i}_s:\forall
i\}$ of $A'$. Then, if
$\rho_s=\sum_ir_i^s\ket{i}_s\bra{i}_s$,
$S(\rho_s)=H(r_i^s)$. Finally, $\forall
i:\enskip
r_i^s=\bra{i}_s\rho_s\ket{i}_s=\Tr\rho_s
\ket{i}_s\bra{i}_s=\Tr\rho_s(P_s^iQ_s)=p_i,
\enskip s=1,2$. Hence,
$$S(\rho_s)=H(p_i)=S(A_s,\rho_{12})\quad
s=1,2.\eqno{(34)}$$

If one performs simultaneous measurement
of $(A_1\otimes 1)$ and of $(1\otimes
A_2)$ on $\rho_{12}$ [denoted by
$(A\wedge A)$], $A_1$ and $A_2$ being the
PTO in the theorem, then one has a
classical discrete joint probability
distribution $p_{ii'}\equiv \Tr
[\rho_{12}(\ket{i}_1\bra{i}_1\otimes
\ket{i'}_2\bra{i'}_2)]$, where $\ket{i}_1
\bra{i}_1=P_1^iQ_1$,
$\ket{i'}_2\bra{i'}_2=P_2^{i'}Q_2$, $A_1=
\sum_ia_iP_1^i+P_1^{\perp}A_1P_1^{\perp}$
(cf (7)), and
$A_2=\sum_{i'}a'_{i'}P_2^{i'}+
P_2^{\perp}A_2P_2^{\perp}$. The
probability distribution implies the
mutual information
$$I(m1:m2)_{A\wedge A}\equiv
H(p_i)+H(p_{i'})-H(p_{ii'})$$ where on
the RHS we have the Shannon entropies
$H(p_{ii'})\equiv
-\sum_{ii'}p_{ii'}lnp_{ii'}$,
$H(p_i)\equiv -\sum_ip_ilnp_i$,
$H(p_{i'}) \equiv
-\sum_{i'}p_{i'}lnp_{i'}$, and $p_i\equiv
\sum_{i'}p_{ii'}$, $p_{i'}\equiv
\sum_ip_{ii'}$ are the marginal
probability distributions.

On account of the crucial PTO property
(5a) (or rather its adjoint), it is
easily seen that $p_{ii'}=\delta_{i,i'}
p_i$. Hence, all three Shannon entropies
equal $H(p_i)=S(A_s,\rho_{12})$, $s=1,2$,
and also $$I(m1:m2)_{A\wedge
A}=S(A_s,\rho_{12})\quad
s=1,2.\eqno{(35)}$$

On the other hand, two chains of
informations are valid:
$$I(m1:m2)_{A\wedge A}\leq
I(m1\rightarrow 2)\leq min\{I(\rho_{12}),
S(\rho_2)\}\eqno{(36a)}$$
$$I(m1:m2)_{A\wedge A}\leq I(1\leftarrow
m2)\leq min\{I(\rho_{12}),S(\rho_1)\}
\eqno{(36b)}$$ (cf (6b) and (7a,b) in
\cite{FH02b}). Here $I(m1\rightarrow 2)$
is the maximal information that one can
gain by measurement on subsystem $1$
about subsystem $2$, and $I(1\leftarrow
m2)$ is the symmetrical quantity. (For a
more precise definition  see
\cite{FH02b}.)

It is seen from (34), (35), and (36a,b)
that we have, what may be called, a
common collapse of the two chains: $$
I(m1:m2)_{A\wedge A}=I(m1\rightarrow 2)=
I(1\leftarrow m2)=S(\rho_2)=S(\rho_1)=
S(A_s,\rho_{12})\quad s=1,2.
\eqno{(37)}$$ Following \cite{Zurek},
\cite{Vedral}, and \cite{FH02b}, we
define
$$I_{qcl}\equiv I(m1\rightarrow 2)=
I(1\leftarrow m2). \eqno{(38)}$$ Hence,
$$I_{qcl}=S(A_s,\rho_{12})\quad s=1,2
\eqno{(39)}$$ as claimed.

Following \cite{Zurek}, we take the
quantum discord $$\delta (\rho_{12})
\equiv I(\rho_{12})-I_{qcl} \eqno{(40)}$$
as the measure of entanglement in
$\rho_{12}$. (In \cite{Vedral} a
different measure of entanglement is
defined. It is independent of $I_{qcl}$.
It coincides with $\delta (\rho_{12})$
for pure states, but not for mixed states
in general.)

In view of (40) and (39), one can
conclude from (33), which has already
been proved, that $$\delta (\rho_{12})=
E_C(A_s,\rho_{12})\quad s=1,2
\eqno{(41)}$$ as claimed.\hfill $\Box$

All pure bipartite states
$\rho_{12}=\ket{\Phi}_{12}\bra{\Phi}_{12}$
are examples to which Theorem 6 applies.
But there are also mixed states of this
kind. To show this we need an auxiliary
result.

{\bf Remark 6.} {\it If
$$\rho_{12}= \sum_kw_k\ket{\Phi
}_{12}^k\bra{\Phi }^k_{12} \eqno{(42)}$$
is any decomposition of a bipartite state
into pure ones, then opposite-subsystem
observables $A_1,A_2$ are PTO for
$\rho_{12}$ {\bf if and only if} they are
PTO for each $\ket{\Phi }_{12}^k$. (The
detectable eigenprojectors $P_s^i$ are
algebraic twin observables, and this
statement has been proved for such
observables, cf C2 in section 3 of
\cite{FHMD}.)}

{\bf Remark 7.} {\it To obtain mixed
states $\rho_{12}$ to which Theorem 6
applies, we define several bipartite pure
states via their Schmidt biorthogonal
decompositions: $$\forall k:\quad
\ket{\Phi}^k_{12}\equiv
\sum_i(r^k_i)^{1/2}\ket{i}_1\ket{i}_2\quad
\forall k,\forall i:\enskip r^k_i>0
\eqno{(43)}$$ (we take $\{r^k_i:\forall
i\}$ distinct for different values of
$k$). Then $A_1\equiv \sum_ia_i
\ket{i}_1\bra{i}_1$ and $A_2\equiv
\sum_ib_i\ket{i}_2\bra{i}_2$ ($a_i$ and
separately $b_i$ distinct) are PTO for
each $\ket{\Phi}^k_{12}$, in particular,
{\bf complete} ones in relation to the
corresponding subsystem states of
$\ket{\Phi}^k_{12}$ (because the
eigenvalues are distinct). Hence, they
are {\bf complete} in relation to the
corresponding subsystem states of
$\rho_{12}$ defined by (42) (cf Remark 6
and Remark 4). Therefore, Theorem 6
applies to this case.}

Let us return to Theorem 5, where
$A_1,A_2$ are PTO but not necessarily
complete in relation to $\rho_1,\rho_2$.
$S(A_s,\rho_{12})\enskip s=1,2$ is
certainly a part of $I_{qcl}$ (as clear
from the argument in the proof of Theorem
6). Hopefully, also
$E_C(A_s,\rho_{12}),\enskip s=1,2$ is a
part of the quantum discord $\delta
(\rho_{12})$.

\section{Conclusion}
Three groups of results have been
obtained in this study.

(i) The concept of physical twin
observables is made more practical by
simplifying it in Theorem 1.

(ii) Utilizing well known relations, a
somewhat surprising general relation
between the coherence entropy $E_C(A,\rho
)$ and the entropy $S(A,\rho )$ of $A$ in
$\rho$ was established (relation (16),
inequality (17), and the Diagram).
Further, it was shown that the coherence
entropy satisfies the third intuitive
requirement (see the Introduction and
Lemma 1). This property led to another
form of  $E_C(A,\rho )$, in which the
redundancies are omitted (Theorem 2).
Finally, it was proved that if one
considers a refinement $A'$ of $A$
instead of the latter, the coherence
entropy cannot decrease. Sufficient and
necessary condition is given for the
increase.

(iii) The case of correlations
incompatibility (Definition 5) comprises
all bipartite pure states and some mixed
ones. The role of the coherence entropy
$E_C(A_s,\rho_{12})$ of the subsystem
observable $A_s$ ($s=1$ or $2$) at issue
in the von Neumann mutual information
$I(\rho_{12})$ is investigated in this
case, and three, more and more specific,
results are obtained:

(a) $I(\rho_{12})$ is the sum of
$E_C(A_s,\rho_{12})$ and a possible
nonnegative term (Theorem 4).

(b) If $A_s$ is one of physical twin
observables in relation to $\rho_{12}$,
then $I(\rho_{12})$ is the sum of the
entropy $S(A_s,\rho_{12})$ of the
observable $A_s$ in $\rho_{12}$ and
$E_C(A_s,\rho_{12})$ and a possible
nonnegative term (Theorem 5).

(c) If the twin observables in (b) are
complete in relation to $\rho_{12}$, then
the third term mentioned in (b) is
necessarily zero (Theorem 6 and relation
(33)). A bipartite mixed-state example is
given (Remark 7).

The result in (c) may be of significance
for the important problem of how to split
$I(\rho_{12})$ into a quasi-classical
part and a part that is the amount of
purely quantum entanglement. Namely, the
term $S(A_s,\rho_{12})$ can be pretty
safely interpreted as precisely the
quasi-classical part. In two references
this was defined \cite{Vedral},
\cite{Zurek}, and in both the entropy
$S(\rho_s)$ of the corresponding
subsystem state $\rho_s$ is an upper
bound for this quantity. On account of
the completeness of the PTO, the entropy
$S(A_s,\rho_{12})$ equals $S(\rho_s)$,
hence, having reached its upper bound, it
must be the quasi-classical part of the
quantum correlations in $\rho_{12}$. The
quantum discord, the term in
$I(\rho_{12})$ that is the excess over
the quasi-classical part, is equal to
$E_C(A_s,\rho_{12})$ in the case
discussed. Thus, the PTO "carry" both the
quasi-classical part (as their entropy in
$\rho_{12}$) and the quantum discord (as
their coherence entropy).\\

{\bf Appendix 1}\\
{\bf Proof} of the equivalence of $PQ=Q$
($Q$ being the range projector of
$\rho$), cf (13), with $\Tr(P\rho )=1$:

{\it Sufficiency:} Since always $\rho
=Q\rho$, one has $\Tr(P\rho )=\Tr[P
(Q\rho )]=\Tr(Q\rho )=\Tr\rho =1$.

{\it Necessity:} Let $\rho
=\sum_nr_n\ket{n}\bra{n}$ be a spectral
form of $\rho$ with positive eigenvalues.
Then, on account of (6), $\Tr(P\rho )=1$
amounts to
$P[\sum_n(r_n\ket{n}\bra{n})]=\sum_n(r_n\ket{n}
\bra{n})$. Multiplying this from the
right by $\ket{n'}\bra{n'}$ with a fixed
$n'$ value, taking the trace and dividing
by $r_{n'}$, one obtains
$\Tr(P\ket{n'}\bra{n'})=1$. Utilizing (6)
again, one has $P\ket{n'}\bra{n'}=
\ket{n'}\bra{n'}$. Finally, since $Q=
\sum_{n'}\ket{n'}\bra{n'}$, the claimed
relation (13) ensues.\hfill $\Box$\\

{\bf Appendix 2}\\
{\bf Proof} of the claim that if the
density operator $\rho$ is nonsingular,
then only the zero event $P=0$ has zero
probability in $\rho$: $$\Tr(P\rho )=0
\enskip \Rightarrow \enskip
\Tr(P^{\perp}\rho )=1 \enskip \Rightarrow
\enskip P^{\perp}Q=Q$$ (cf Appendix 1).
But now $Q=1$. Hence, $P^{\perp}=1$, and
$P=0$.\hfill $\Box$

\end{document}